Merrifield's Variational Ansatz is extended so as to cover the case of two electronic bands mixed by an Einstein phonon. The Hamiltonian is composed of the local and kinetic (hopping) energies in the absence of vibrations, the vibrational energy, and a mixing band-off-diagonal part linear in the electron-phonon coupling, all expressed in second quantization terms. The variational eigenstate is a linear combination of Merrifield states for either electronic band. We derive equations for the phonon amplitudes in momentum space and for the fractional contribution of either electronic band to the overall variational state.

1. Introduction

The concept of 'Polaron in a Solid' is still enigmatic to some extent, now that more than 60 years have passed since the term has been introduced in the Physics Vocabulary (see Pekar [1]). A powerful theoretical method for studying the polaron (charge carrier & associated lattice distortion) is based on defining an appropriate variational eigenstate comprising fermion and boson ladder (creation and annihilation) operators to diagonalize the polaron Hamiltonian and find the minimal (ground state) energy. A number of good proposals have been made ever since the pioneering work by Lee, Low, and Pines (see Appel [2]) which include contributions by Merrifield [3], Toyozawa [4] and Davidov [5] to mention a few (for a review see Lindenberg, Zhao, and Brown [6]). Lately, more precise variational techniques have been developed by the Global Local Method [7] which gives most accurately the ground state energy of the Holstein polaron.

The essential point in a variational Ansatz is the way it defines the structure of the phonon cloud surrounding the charge carrier through the "phonon form factor". In this respect, Merrifield's approach [3] which extends the earlier proposal by Lee, Low, and Pines by inroducing field operators for the electron too, is perhaps most transparent physically and leads to tractable results. Although originally aimed at Holstein polarons (excitons), which form through the interaction of an electronic carrier with symmetry-retaining phonons along a linear chain, it should be extendable to lower symmetries through incorporating the vibronic mixing effects.

A variational Ansatz has earlier been applied to describe an itinerant lower-symmetry Jahn-Teller (JT) polaron in 1D.[8] Two degenerate narrow electron bands have been considered with their specific hopping energies. The coupling term has been assumed of the band-diagonal type in which the phonon field couples to the difference in electron density of the two electronic bands. However, this type of interaction Hamiltonian failing to provide any genuine mixing transitions between the two electronic bands, it is discarded by other authors as being non-vibronic [9]. One way or the other, an energy versus total momentum relation has been computed numerically for any of the constituent bands and shown to be polaronic in character.

Our Merrifield-based analysis indicates that in case of a band-diagonal coupling to the difference in electron density as above the system collapses into a single-band Holstein polaron in either the first or the second constituent band with corresponding phonon amplitudes out-of-phase to each other. In general, there appear to be two ways of splitting an electronic degeneracy in collective Jahn-Teller phenomena:

(i) A genuine vibronic (band off-diagonal) mixing by a symmetry-breaking vibrational mode resulting in a new lower-symmetry phase;

(ii) A non-mixing (band-diagonal) coupling to a symmetry-retaining mode resulting in the same symmetry configuration. The energy bands split in this case because the displaced-oscillator energy enters with different signs for the two bands (that is, it adds up and subtracts, respectively).

It therefore appears that no variational study of the genuine vibronic polaron has so far been made and puts an even stronger impetus on our present effort.

There has been an increasing appreciation lately of the possible role of vibronic polarons in the electric transport of transition-metal compounds [10]. As a matter of fact, the search for Jahn-Teller polarons in $La_{2-x}Sr_xCuO_4$ has led to the discovery of a high-$T_c$ superconductivity [11]. Similar Jahn-Teller polarons are currently considered with regard to the observed colossal magnetoresistance in $La_{1-x}Ca_xMnO_3$ and related materials [12]. On the other hand, Pseudo-Jahn-Teller (PJT) polarons forming as fermionic excitations scatter from their self-consistently associated double wells are often mentioned as the principal charge carriers maintaining an axial leak which effects the interplane coupling in single-layer cuprates, such as $La_{2-x}Sr_xCuO_4$ [13].

In a preceding Part I we described the general conditions to be met while building up a variational solution for the vibronic polarons [14]. The present Part II is organized as follows: We first define a Hamiltonian for the vibronic polaron along a linear chain. We construct a variational eigenstate as a linear combination of two Merrifield band states and derive the matrix elements of the total Hamiltonian (diagonal and off-diagonal). These matrix elements are used for bilding up variational equations for the phonon amplitudes in momentum space, as well as for the fractional amplitudes of the two band eigenstates. Symmetry requirements are imposed on the amplitudes before the variational equations are finally prepared for the numerical calculations in Part III. Some analytic conclusions will also be made beforehand wherever possible.

## 2. Vibronic Hamiltonian

The vibronic polaron case requires the availability of two narrow (nearly-) degenerate electronic bands and a mixing phonon field of the appropriate symmetry.(See Bersuker [15-17], Bersuker and Polinger [18]). In so far as the band off-diagonal coupling may be expected to break the original site symmetry, the mixing mode will have to be one of the symmetry-breaking vibrations transforming according to the irreducible representations (or their direct sums) of the point group at that site. In the particular case of the latter group containing the spatial inversion, the mixing mode will be odd parity if the respective electronic bands compose of states of the opposite parities. For instance, if the corresponding bands form of 3d and 4p axial orbitals, even- and odd-parity, respectively, the mixing vibration may be one of the $A_{2u}$ or the $E_u$ odd-parity modes, both being symmetry-breaking in their nature. For sufficiently strong coupling, the octahedral (tetragonal) site symmetry will be lowered to that of an $A_{2u}$- or $E_u$- deformed octahedron.

The vibronic Hamiltonian reads (in second quantization terms):

$$H_{vib} \equiv H_0 + H_{kin} + H_{int} + H_{ph} = \sum_{n,\mu} \varepsilon_{n,\mu} a_{n,\mu}^\dagger a_{n,\mu} + \sum_{n,\mu} j_{n,\mu} (a_{n,\mu}^\dagger a_{n+1\mu} + a_{n-1,\mu}) +$$

$$\sum_{n,\mu,\nu} g^{n,\mu,\nu} a_{n,\mu}^\dagger a_{n,\nu} (b_n^\dagger + b_n) + \sum_n \eta\omega_n b_n^\dagger b_n, \qquad (1)$$

where $H_0$, $H_{kin}$, $H_{int}$, and $H_{ph}$ denote the four sums in the order of their appearance: the local energy, the kinetic (hopping) energy, the electron-phonon interaction energy, and the phonon energy, respectively. Also n is the site label, $\mu$ and $\nu$ are band labels ( $\mu,\nu = 1,2$ ). $a_{n,\mu}^\dagger$ ($a_{n,\mu}$) are fermion creation (annihilation) operators, $b_n^\dagger$ ($b_n$) are boson creation (annihilation) operators, $\varepsilon_{n,\mu}$ are the local fermion energies, $j_{n,\mu}$ are the fermion hopping energies, $g^{n,\mu,\nu}$ are the fermion-boson coupling constants, $\omega_n$ are the phonon frequencies. Point-group symmetry requires that the representation for the phonon field be included in the direct product of the representations for the two electronic bands: $\Gamma_{ph} \subset \Gamma_1 \otimes \Gamma_2$.

Regarding the PJT mixing of definite-parity components, $\mu$ and $\nu$ label electronic bands composed of states of the opposite parities, even and odd, respectively, $b_n^\dagger$ create odd-parity phonons. Assuming the atomic chain to be isotropic, we take the underlying parameters $\varepsilon_{n,\mu} = \varepsilon_\mu / N$ through $g^{n,\mu,\nu} = g^{\mu\nu} / N$ out of the sums, where N is the number of unit cells. Doing likewise with $\omega$ implies adopting an Einstein model for the mixing phonon. Intrinsic PJT mixing obtains for $g^{\mu\mu} = g^{\nu\nu} = 0$.

So far $H_{vib}$ has only been diagonalized at some particular values of the ingredient parameters, namely at $j_\mu = 0$ or at $g^{\mu\nu} = 0$. The former describes a local dynamic picture which has been studied within the adiabatic approximation (see [19,20]). Reproducing the essentials, we shall make use of the adiabatic results as guidelines for checking the variational conclusions at low hopping energies.

### 2.1. Local dynamics

In the adiabatic approximation, the electron-phonon coupling term is dealt with semi-classically through introducing a mode coordinate $Q = \sqrt{\{\eta\omega / K\}} \times (b_n^\dagger + b_n)$ regarded as a c-number. Accordingly, the local Hamiltonian at $j_\mu = 0$ turns into:

$$H_{vib}^{local} = \tfrac{1}{2} e_{12} (|2\rangle\langle 2| - |1\rangle\langle 1|) + GQ (|2\rangle\langle 1| + |1\rangle\langle 2|) + \tfrac{1}{2} ( \mathbf{P}^2 / M + KQ^2) \qquad (2)$$

Here and above, $K = M\omega^2$ is the stiffness, M is the reduced mass of the vibrator, $G = g^{\mu\nu} \sqrt{(K / \eta\omega)}$ is the local electron-mode coupling constant, and $e_{12} = |\varepsilon_2 - \varepsilon_1|$ is the interlevel energy gap. In the basis of the ($|1\rangle$, $|2\rangle$) eigenstates of the electronic part, the first-order perturbation local adiabatic energies are:

$$E_\pm(Q) = \tfrac{1}{2} \{ KQ^2 \pm \sqrt{[ (2GQ)^2 + e_{12}^2]} \} \qquad (3)$$

The upper adiabatic branch ($E_+$) is always minimal at $Q = 0$, while the lower one ($E_-$), also minimal at $Q = 0$ for $e_{12} > 4\, e_{JT}$, develops an instability (lateral minimums) at $\pm Q_0$, as the extremum at $Q = 0$ turns into a maximum in between at $e_{12} < 4\, e_{JT}$. Here

$$e_{JT} = G^2 / 2K \qquad (4)$$

is the Jahn-Teller energy and

$$Q_0 = \sqrt{[(2\, e_{JT} / K)(1 - \eta^2)]}, \qquad (5)$$

with $\eta = e_{12} / 4e_{JT}$, is the distorted-configuration coordinate. The essential point is that the local polaron which forms at $\eta < 1$ is off-centered relative to $Q = 0$, in the sense that its configurational environment exhibits the lower symmetry at $Q_0$. It is worth seeing whether this symmetry feature appearing on the local lower-energy profile can withstand the opposite increased-energy trends of the finite $j_\mu$, thereby providing stability for the vibronic polaron.

### 3. Variational eigenstate

We compose a variational eigenstate as a linear combination of Merrifield-type states for either of the constituent bands:

$$|\psi(\kappa)\rangle = N^{-1/2} \sum_n \exp(i\kappa n) |\phi_n^\kappa\rangle \qquad (6)$$

with

$$|\phi_n^\kappa\rangle = \sum_\mu \alpha_\mu^\kappa\, a_{n\mu}^\dagger \exp\{-N^{-1/2} \sum_q [\beta_{q\mu}^\kappa \exp(-iqn)\, b_q^\dagger - \beta_{q\mu}^{\kappa *} \exp(+iqn)\, b_q]\} |0\rangle \qquad (7)$$

An equivalent though useful reformulation reads:

$$|\psi(\kappa)\rangle = \sum_\mu \alpha_\mu^\kappa |\psi_\mu(\kappa)\rangle \qquad (8)$$

with

$$|\psi_\mu(\kappa)\rangle = N^{-1/2} \sum_n \exp(i\kappa n)\, a_{n\mu}^\dagger \exp\{-N^{-1/2} \sum_q [\beta_{q\mu}^\kappa \exp(-iqn)\, b_q^\dagger - \beta_{q\mu}^{\kappa *} \exp(+iqn) b_q]\} |0\rangle \qquad (9)$$

Here and above $\alpha_\mu$ stand for the fractional band amplitudes. The single-band Merrifield eigenstates are normalized automatically. Consequently, the normalization condition for their linear combination reads:

$$\langle \psi(\kappa)|\psi(\kappa)\rangle = N^{-1} \sum_{n,m} \exp\{i(n-m)\kappa\} \langle \phi_m^\kappa | \phi_n^\kappa \rangle = \sum_\mu |\alpha_\mu^\kappa|^2 \qquad (10)$$

For deriving the variational equations, it would be simpler to define a normalized state of the form:

$$|\Psi(\kappa)\rangle = |\psi(\kappa)\rangle / \langle \psi(\kappa)|\psi(\kappa)\rangle^{1/2} \qquad (11)$$

in which the factor $<\psi(\kappa)|\psi(\kappa)>$ bears no explicit dependence on the phonon amplitudes, in so far as the right-hand side of eq.(10) is set equal to 1 as a "rigid normalization condition."

Merrifield's eigenstates have been discussed elsewhere in the literature [3,21]. They describe the phonon structure in terms of coherent vibrational states. This method extends the early proposal by Lee, Low and Pines by introducing field operators for the fermion carriers as well. The chief merit of Merrifield's approach seems to be in that it leads to physically-transparent conclusions.

## 4. Expectation value

Using the normalized variational state, we derive the expectation value of the Hamiltonian:

$$<\Psi|H|\Psi> \ = \ <\psi|\psi>^{-1} \left( \sum_\mu |\alpha_\mu^\kappa|^2 \varepsilon_\mu + \sum_\mu |\alpha_\mu^\kappa|^2 j_\mu \{ \exp(i\kappa) S_{+1}^{\mu\mu} + \exp(-i\kappa) S_{-1}^{\mu\mu} \} + \right.$$

$$\left. N^{-1} \eta\omega \sum_{q\mu} |\alpha_\mu^\kappa|^2 |\beta_{q\mu}^\kappa|^2 + N^{-1} \sum_{q\mu\nu} \alpha_\mu^{\kappa*} \alpha_\nu^\kappa g^{\mu\nu} Q_{\mu\nu} S_{0\kappa}^{\nu\mu} \right), \tag{12}$$

where the following notations have been introduced:

$$S_{\pm 1\kappa}^{\mu\mu} = \exp\{-(1/N) \sum_q |\beta_{q\mu}^\kappa|^2 [1 - \exp(\mu i q)]\} \tag{13}$$

$$S_{0\kappa}^{\mu\nu} = \exp\{-(1/2N) \sum_q [\,|\beta_{q\mu}^\kappa|^2 + |\beta_{q\nu}^\kappa|^2 - 2\beta_{q\mu}^{\kappa*} \beta_{q\nu}^\kappa\,]\} \tag{14}$$

$$Q_{\mu\nu}^\kappa = (1/N) \sum_q ( \beta_{-q\mu}^{\kappa*} + \beta_{q\nu}^\kappa ) \tag{15}$$

Each of these quantities has a specific physical meaning: $S_{\pm 1\kappa}^{\mu\mu}$ and $S_{0\kappa}^{\mu\nu}$ are Debye-Waller (DW) factors, band-diagonal and off-diagonal, respectively, $Q_{\mu\nu}^\kappa$ for $\mu \neq \nu$ is a replica of the real-space mixing-mode coordinate.

It should be noted that the same averages are obtained irrespective of whether the electronic operators are subjected to commutation or anticommutation. In this respect the present method applies equally well to (fermionic) electronic polarons and to (nearly-bosonic) excitonic polarons.

## 5. Variational equations

The present problem is one of the following set of variational parameters: $\beta_{q1}^\kappa$, $\beta_{q2}^\kappa$ $\alpha_1^\kappa$ and $\alpha_2^\kappa$, each set for any specific value of the total crystalline momentum $\kappa$.

From the derivative $\partial/\partial \beta_{q\mu}^{\kappa*} <\Psi|H|\Psi> = 0$, we obtain the following system of linear equations for the two sets of phonon amplitudes, regarding the Debye-Waller factors, phases, and mode coordinates as β-independent quantities (this will lead to self-consistent equations for the β's):

$$\beta_{q\mu}^\kappa = -\{ g^{\mu\mu} |\alpha_\mu^\kappa|^2 + g^{\mu\nu} \alpha_\nu^\kappa \alpha_\mu^{\kappa*} S_{0\kappa}^{\nu\mu} + g^{\nu\mu} \alpha_\nu^\kappa \alpha_\mu^{\kappa*} Q_{\nu\mu} S_{0\kappa}^{\nu\mu} \beta_{qv}^\kappa \} /$$

$$\{ \eta\omega |\alpha_\mu^\kappa|^2 [1 + 4(j_\mu/\eta\omega) S_\kappa^{\mu\mu} \sin(\kappa - \Phi_\kappa^{\mu\mu} - q/2) \sin(q/2)] -$$

$$\tfrac{1}{2} [ g^{\mu\nu} \alpha_\nu^{\kappa*} \alpha_\mu^\kappa Q_{\mu\nu}^\kappa S_{0\kappa}^{\mu\nu} + g^{\nu\mu} \alpha_\nu^\kappa \alpha_\mu^\kappa Q_{\nu\mu} S_{0\kappa}^{\nu\mu} ] \} \tag{16}$$

These are readily solved and implying $\mu = 1$, $\nu = 2$ we get ($g^{\mu\nu} = g^{\nu\mu}$ assumed throughout):

$$\beta_{q\mu}^{\kappa} = \{(g^{\mu\mu}/\eta\omega) + (g^{\mu\nu}/\eta\omega)(\alpha_{\nu}^{\kappa}/\alpha_{\mu}^{\kappa})S_{0\kappa}^{\nu\mu}\}\{-D_{q\nu\mu}^{\kappa} + (g^{\nu\mu}/\eta\omega)(\alpha_{\mu}^{\kappa}/\alpha_{\nu}^{\kappa}) S_{0\kappa}^{\mu\nu} Q_{\nu\mu}^{\kappa}\}/D_q \qquad (17)$$

$$\beta_{q\nu}^{\kappa} = \{(g^{\nu\nu}/\eta\omega) + (g^{\nu\mu}/\eta\omega)(\alpha_{\mu}^{\kappa}/\alpha_{\nu}^{\kappa}) S_{0\kappa}^{\mu\nu}\}\{-D_{q\mu\nu}^{\kappa} + (g^{\mu\nu}/\eta\omega) (\alpha_{\nu}^{\kappa}/\alpha_{\mu}^{\kappa}) S_{0\kappa}^{\nu\mu} Q_{\mu\nu}^{\kappa}\}/D_q \qquad (18)$$

The following symbols have been used:

$$D_{q\mu\nu}^{\kappa} = 1 + 4(j_{\mu}/\eta\omega)S_{\kappa}^{\mu\mu} \sin(\kappa - \Phi_{\mu\mu}^{\kappa} - q/2)\sin(q/2) - (g^{\mu\nu}/\eta\omega)\text{Re}[(\alpha_{\nu}^{\kappa}/\alpha_{\mu}^{\kappa})Q_{\nu\mu}^{\kappa} S_{0\kappa}^{\nu\mu}] \qquad (19)$$

$$D_q = D_{q\mu\nu}^{\kappa} D_{q\nu\mu}^{\kappa} - (g^{\mu\nu}/\eta\omega)^2 |S_{0\kappa}^{\mu\nu}|^2 |Q_{\mu\nu}^{\kappa}|^2 \qquad (20)$$

$$S_{\mu\mu}^{\kappa} = \exp\{-(1/N)\Sigma_q|\beta_{q\mu}^{\kappa}|^2[1 - \cos(q)]\} \qquad (21)$$

$$\Phi_{\mu\mu}^{\kappa} = (1/N) \Sigma_q|\beta_{q\mu}^{\kappa}|^2 \sin(q) \qquad (22)$$

The two equations for $\beta_{q\mu}^{\kappa}$ and $\beta_{q\nu}^{\kappa}$ are reproduced explicitly, even though they obtain from the same common formula by interposing the two band indices.

For the variation in, say $\alpha_{\mu}^{\kappa*}$, we present the energy functional in a convenient form:

$$E(\kappa) \equiv \langle \Psi(\kappa)| H |\Psi(\kappa)\rangle = (\langle \psi | \psi \rangle)^{-1} \Sigma_{\nu\mu} \alpha_{\mu}^{\kappa} \alpha_{\nu}^{\kappa*} \varepsilon_{\mu\nu}^{\kappa} \qquad (23)$$

where with all the couplings included

$$\varepsilon_{\mu\mu}^{\kappa} \equiv \varepsilon_{\mu} + j_{\mu} \{e^{i\kappa}S_{+1\kappa}^{\mu\mu} + e^{-i\kappa}S_{-1\kappa}^{\mu\mu}\} + (\eta\omega) (1/N)\Sigma_q| \beta_{q\mu}^{\kappa} |^2 + g^{\mu\mu} Q_{\mu\mu}^{\kappa}$$

$$= \varepsilon_{\mu} + 2 j_{\mu} S_{\mu\mu}^{\kappa} \cos(\kappa - \Phi_{\mu\mu}^{\kappa}) + g^{\mu\mu} Q_{\mu\mu}^{\kappa} + \eta\omega (1/N) \Sigma_q | \beta_{q\mu}^{\kappa} |^2 \qquad (24)$$

$$\varepsilon_{\mu\nu}^{\kappa} = g^{\mu\nu} Q_{\mu\nu}^{\kappa} S_{0\kappa}^{\mu\nu} \qquad (25)$$

Now, the $\alpha$-derivatives yield another pair of extremal conditions:

$$\delta_{\alpha\nu}^{\kappa*} E(\kappa) = \{1 / (\langle \psi | \psi \rangle)^2\} \{ \Sigma_{\mu\nu'} \alpha_{\nu}^{\kappa} \varepsilon_{\mu\nu'}^{\kappa} \delta_{\nu\nu'} \langle \psi | \psi \rangle - \alpha_{\nu}^{\kappa} \Sigma_{\mu\nu'} \alpha_{\nu'}^{\kappa*} \alpha_{\mu}^{\kappa} \varepsilon_{\mu\nu'}^{\kappa} \} = 0 \qquad (26)$$

which turns into a simple relation:

$$\alpha_{\nu}^{\kappa} = E(\kappa)^{-1} \Sigma_{\mu} \alpha_{\mu}^{\kappa} \varepsilon_{\nu\mu}^{\kappa} \qquad (27)$$

under the normalization condition

$$\Sigma_{\mu} | \alpha_{\mu}^{\kappa} |^2 = 1 \qquad (28)$$

In the absence of an interband mixing ($g^{\mu\nu} = 0$), the two-band system collapses into a single-band Holstein problem, as verified easily by inspecting the α-variational equations. We further retain the interband coupling mainly by setting $g^{\mu\mu} = g^{\nu\nu} = 0$. Nevertheless, the general solution comprising both the intra-and inter- band couplings is interesting in itself and is worth a complete investigation.

From the α-variational conditions (27) written separately for μ and ν, we get the following system of homogeneous equations:

$$\alpha_\mu^\kappa = \alpha_\nu^\kappa \varepsilon_{\mu\nu}^\kappa / ( \sum_{\mu'\nu'} \alpha_{\mu'}^\kappa \alpha_{\nu'}^{\kappa*} \varepsilon_{\mu'\nu'}^\kappa - \varepsilon_{\mu\mu}^\nu) \tag{29}$$

$$\alpha_\nu^\kappa = \alpha_\mu^\kappa \varepsilon_{\nu\mu}^\kappa / ( \sum_{\mu'\nu'} \alpha_{\mu'}^\kappa \alpha_{\nu'}^{\kappa*} \varepsilon_{\mu'\nu'}^\kappa - \varepsilon_{\nu\nu}^\kappa) \tag{30}$$

For this system to have a non-vanishing solution, its determinant must be zero where:

$$\text{Det} = [E(\kappa)^2 - (\varepsilon_{\mu\mu}^\kappa + \varepsilon_{\nu\nu}^\kappa)E(\kappa) + \varepsilon_{\mu\mu}^\kappa \varepsilon_{\nu\nu}^\kappa - \varepsilon_{\mu\nu}^\kappa \varepsilon_{\nu\mu}^\kappa] / \prod_{\mu'}(E(\kappa) - \varepsilon_{\mu'\mu'}^\kappa) \tag{31}$$

Unless $E(\kappa) = \varepsilon_{\mu'\mu'}^\kappa$, the resulting equation for $E(\kappa)$ has two roots reading:

$$E(\kappa)_\pm = \tfrac{1}{2} \{ (\varepsilon_{\mu\mu}^\kappa + \varepsilon_{\nu\nu}^\kappa) \pm \sqrt{[(\varepsilon_{\mu\mu}^\kappa - \varepsilon_{\nu\nu}^\kappa)^2 + 4\,\varepsilon_{\mu\mu}^\kappa \varepsilon_{\nu\nu}^\kappa]}\} \tag{32}$$

Formally, these roots are the first-order perturbation eigen-energies of the complete Hamiltonian in the $(\psi_\mu^\kappa, \psi_\nu^\kappa)$ basis.

At the same time, however, the extremal variational energy being

$$E(\kappa) = \sum_{\mu'\nu'} \alpha_{\mu'}^\kappa \alpha_{\nu'}^{\kappa*} \varepsilon_{\mu'\nu'}^\kappa \tag{33}$$

it raises the legitimate question as to whether $E(\kappa)$ and $E(\kappa)_\pm$ agree with each other. Pursuing the matter, from

$$\alpha_\nu^\kappa = \alpha_\mu^\kappa \varepsilon_{\nu\mu}^\kappa / [ E(\kappa)_\pm - \varepsilon_{\nu\nu}^\kappa ] \tag{34}$$

$$\sum_{\mu'} |\alpha_{\mu'}^\kappa|^2 = 1 \tag{35}$$

we readily derive

$$|\alpha_\mu^\kappa| = [1 + |\varepsilon_{\nu\mu}^\kappa|^2 / |E(\kappa)_\pm - \varepsilon_{\nu\nu}^\kappa|^2]^{-1/2} \tag{36}$$

$$|\alpha_\nu^\kappa| = [1 + |\varepsilon_{\mu\nu}^\kappa|^2 / |E(\kappa)_\pm - \varepsilon_{\nu\nu}^\kappa|^2]^{-1/2} \tag{37}$$

Next, substituting for the α's in

$$E(\kappa) = |\alpha_\mu^\kappa|^2 \varepsilon_{\mu\mu}^\kappa + |\alpha_\nu^\kappa|^2 \varepsilon_{\nu\nu} + 2\mathrm{Re}(\alpha_\mu^\kappa \alpha_\nu^{\kappa*} \varepsilon_{\nu\mu}^\kappa) \tag{38}$$

we get after some lengthy manipulations:

$$E(\kappa) = E(\kappa)_- \tag{39}$$

i.e. the extremal (ground-state) energy coincides with the lower perturbation-energy branch.

## 6. Iteration starting conditions

The subsequent two Sections will deal with setting the starting conditions of an iteration procedure for solving the variational equations numerically. Two sets will be considered including what we call the "Small Polaron" and "Large Polaron" extremes.

### 6.1. Small polaron extreme

A small polaron solution will be composed of phonon amplitudes which have finite components all over the momentum space. Assuming the Fourier amplitudes $\beta_{q\mu}^{\kappa}$, etc. to be slowly varying with q and introducing the average number of phonons in either band:

$$N_\mu^\kappa = (1/N) \sum_q |\beta_{q\mu}^\kappa|^2 \tag{40}$$

we get the following "small polaron" quantities:

$$S_{\mu\mu}^\kappa = \exp(-N_\mu^\kappa) \tag{41}$$

$$S_{0\kappa}^{\mu\mu} = \exp[-(1/2N) \sum_q (|\beta_{q\mu}^\kappa| - |\beta_{qv}^\kappa|)^2] \tag{42}$$

$$\Phi_{\mu\mu}^\kappa = 0 \tag{43}$$

$$Q_{\mu\mu}^\kappa = (1/N) \sum_q (|\beta_{q\mu}^\kappa| + |\beta_{-qv}^{\kappa*}|) \exp\{i\phi q\} \tag{44}$$

provided the respective phonon components in momentum space are in-phase for the two bands (making the off-diagonal DW factor real): $\beta_\mu^\kappa = |\beta_\mu^\kappa| \exp\{i\phi_q^\kappa\}$, etc. Inserting into the extremal energy equation, we get a "small-polaron ground-state energy":

$$E(\kappa) = \tfrac{1}{2} \{2(j_\mu^\kappa \exp(-N_\mu^\kappa) + j_v^\kappa \exp(-N_v^\kappa))\cos(\kappa) + (\eta\omega)(N_\mu^\kappa + N_v^\kappa) \pm [\{e_{\mu v} + 2(j_\mu^\kappa \exp\{-N_\mu^\kappa\} -$$

$$j_v^\kappa \exp\{-N_v^\kappa\})\cos(\kappa) + (\eta\omega)(N_\mu^\kappa - N_v^\kappa)]^2 + 4(g^{\mu v})^2 |S_{0\kappa}^{\mu v}|^2 |Q_{\mu v}^\kappa|^2\}^{1/2} \} \tag{45}$$

where $e_{\mu v} = |\varepsilon_\mu - \varepsilon_v|$. We see that the hopping terms each renormalize with the respective number-of-phonon exponentials (band-diagonal DW factors). The mixing energy does likewise with the off-diagonal DW factor but in addition it depends on the mixing-mode coordinate.

When the phonon occupation numbers $N_\mu^\kappa$ are large, the small polaron is immobilized by a heavy translational mass, its E(κ) dependence becoming flatter. There are now all the prerequisites for applying the local adiabatic result of Section 2.1 where the mixing energy is $2GQ_0$ at the small polaron configuration. Quantizing the adiabatic coordinate, one obtains:

$$g^{\mu\nu} < \mu | (b_n^\dagger + b_n) | \nu > = g^{\mu\nu} | Q_{\mu\nu}^{\kappa} | \qquad (46)$$

Clearly, the adiabatic approach overestimates the off-diagonal DW factor which is unity only when the phonon occupation numbers are the same in both component bands. Comparing with the semiclassical adiabatic mixing-energy term:

$$g^{\mu\nu} \sqrt{\{2e_{JT} / \eta\omega\}} \sqrt{\{1 - (e_{\mu\nu}/4e_{JT})^2\}} \qquad (47)$$

we get

$$| Q_{\mu\nu}^{\kappa} | = \sqrt{(2e_{JT} / \eta\omega)} \sqrt{[1 - (e_{\mu\nu}/4e_{JT})^2]} \qquad (48)$$

where for adiabatic small polarons $2e_{JT} \gg \eta\omega$, $4e_{JT} \gg |e_{\mu\nu}|$.

### 6.2. Large polaron extreme

We will define the "Large Polaron" as the one which builds up by phonon amplitudes from a highly localized region in q-space. Accordingly, we set $\beta_{q\mu}^{\kappa} = \beta_{\mu}^{\kappa} N \delta(q)$. The "large polaron phonon coordinate" then obtains as:

$$Q_{\mu\nu}^{\kappa} = \beta_{q\mu}^{\kappa} + \beta_{q\nu}^{\kappa} \qquad (49)$$

From the adiabatic weak-coupling theory, the large-polaron coordinate is

$$\sim \sqrt{\{|e_{\mu\nu}| / 2\eta\omega\}} \sqrt{\{1 - (e_{\mu\nu}/4e_{JT})^2\}} \qquad (50)$$

where $4e_{JT} \sim |e_{\mu\nu}|$. Comparing with $Q_{\mu\nu}^{\kappa}$, we get an estimate for the phonon amplitudes to start the iteration with.

### 7. Symmetry requirements

The general symmetry requirement for the interaction Hamiltonian is that it must remain invariant under the operations of the point group. This requirement will be met if the representation for the mixing mode is a subset of the direct product of the representations transforming the two electronic states to be mixed: $\Gamma_{vib} \subset \Gamma_1 \otimes \Gamma_2$.

### 7.1. Pseudo-Jahn-Teller mixing

The Pseudo-Jahn-Teller effect is the mixing of two nearly-degenerate electronic states by a phonon mode of the appropriate symmetry. If the two electronic states to be mixed are of definite parities and if they are opposite-parity, then the mixing vibration should be odd-parity to meet the group theoretical requirement.

The symmetry of the linear chain under consideration may be expected to impose additional though general conditions on the phonon amplitudes. The point groups for an infinite chain of atoms are well known: $C_{\infty v}$ and $D_{\infty h}$ comprising reflections from vertical and horizontal planes, respectively.

The second one has two types of odd-parity symmetry-breaking normal vibrations: longitudinal $\Sigma$ and transverse $\Pi$ with respect to the chain axis, namely $\Sigma_u^{\pm}$ and $\Pi_u$, respectively. The sign $\pm$ in the notation for the former tells of whether the character of the reflection from a vertical plane $\infty\,\sigma_v$ is $\pm 1$, respectively. Of these, the positive character mode $\Sigma_u^{+}$ is of interest since it transforms as the coordinate z along the chain axis. The other mode $\Pi_u$ comprises transverse coordinates which transform as (x,y), perpendicular to the chain.

Undoubtedly, our 1D linear-chain method is consistent with the coupling to a $\Sigma_u^{+}$ vibration of the infinite-chain point group. These longitudinal displacements should be regarded as ones of the unit-cell center of gravity relative to its equilibrium position which displacements induce an electric dipole moment within the unit cell, due to their asymmetric character.

For instance, if we consider a linear sub-chain composed of a transition metal atom in the midst of two oxygen atoms on its both sides to represent a unit cel, then the in-phase oxygen vibrations along the chain axis will displace the center of gravity of the O-M-O di-oxygen frame from left to right across the site of the metal atom M which may remain immobile in this mode.

### 7.2. Dynamic Jahn-Teller effect

The Dynamic-Jahn-Teller effect involves the vibronic mixing of two degenerate electronic states, while taking account of the coupling between these states. In as much as the two electronic states to be mixed are of the same parity, the mixing mode should transform according to a symmetry-breaking even-parity irreducible representation of the point group.

For an infinite linear chain, we have $\Sigma_g \otimes \Sigma_g = \Sigma_g \oplus A_1$ and $\Sigma_u^{\pm} \oplus \Sigma_u^{\pm} = \Sigma_g \oplus A_1$, where $A_1$ is the fully symmetric (breathing mode) representation of the point group. We see that the symmetric quadrate of the longitudinal mode of either parity decomposes into its even-parity counterpart and the breathing mode. For this reason, a thorough consideration of the point group requires introducing two coordinates, $Q_\Sigma$ and $Q_A$, respectively, and the complete interaction Hamiltonian is:

$$H_{int} = \sum_n g_\Sigma (a_{n2}^\dagger a_{n1} + a_{n1}^\dagger a_{n2}) Q_\Sigma + \sum_n g_A (a_{n2}^\dagger a_{n2} - a_{n1}^\dagger a_{n1}) Q_A \qquad (51)$$

While most of the authors have considered the effect of $Q_\Sigma$ alone, which is the symmetry-breaking vibronic-mixing coordinate, a few have only introduced $Q_A$, which is the symmetry-retaining coordinate. However, in the latter case, the system logically collapses into a symmetry-retaining Holstein polaron in any of the constituent bands so that the vibronic-mixing effect is overlooked.

### 7.3. Example: The O-M-O triatomic molecule

The basic element of the $MO_6$ octahedron, a central building block of both the cuprates and the manganates, is the O(A)-M(P)-O(A) triatomic molecule where A stands for "apical" and P for "planar". The energy spectrum of that molecule is well-known. It consists of a "bonding" (B), "non-bonding" (NB), and "antibonding" (AB) energy levels, as follows (cf. [22]):

$$E_B = E_O + \tfrac{1}{2} \{ \varepsilon_{O-M} - \sqrt{[(\varepsilon_{O-M})^2 + 8(t_{M-O})^2]} \} \qquad (52)$$

$$E_{NB} = E_O \tag{53}$$

$$E_{AB} = E_O + \tfrac{1}{2}\varepsilon_{O-M} + \sqrt{\{(\varepsilon_{O-M})^2 + 8(t_{M-O})^2]\}} \tag{54}$$

in the increasing-energy order. Here $\varepsilon_{O-M} = |\varepsilon_M - \varepsilon_O|$ where $\varepsilon_O$ and $\varepsilon_M$ are the energies of the oxygen $|2p_z\rangle$ and metal $|3d_{z^2}\rangle$ orbitals, respectively, when these ions are far apart, $t_{M-O}$ is the hybridization (M-O hopping) energy. $E_O$ is a mostly oxygen level which is doubly-degenerate in the absence of hybridization with the metal orbitals. For $t_{M-O} \ll \varepsilon_{O-M}$, the hybridization brings about but a small splitting of the doublet levels.

It may be instructive to reproduce the eigenstates corresponding to the above eigenvalues:

$$\Psi_O = (1/\sqrt{2})(|1\rangle + |2\rangle) \tag{55}$$

$$\Psi_B = \sin(\gamma)|0\rangle + (1/\sqrt{2}) \times \cos(\gamma)(|1\rangle - |2\rangle) \tag{56}$$

$$\Psi_{AB} = \sin(\beta)|0\rangle + (1/\sqrt{2}) \times \cos(\beta)(|1\rangle - |2\rangle) \tag{57}$$

where $|0\rangle$ stands for the $3d_{z^2}$ orbital state, while $|1\rangle$ and $|2\rangle$ are the $|2p_z\rangle$ orbitals of the two lateral oxygens. Here

$$\beta = \arctan(E_{AB}/t_{M-O}\sqrt{2}) \tag{58}$$

$$\gamma = \arctan(E_B/t_{M-O}\sqrt{2}) \tag{59}$$

The symmetry group of the $MO_6$ octahedron is $O_h$ in the absence of distortion. The symmetry-breaking vibrations of the O(A)-M(P)-O(A) unit involve oxygen displacements mainly, due to the smaller oscillator mass. These are the even parity $E_g$ and $T_{2g}$ and odd parity $T_{1u}$ and $T_{2u}$. $T_{1u}$ splits to planar- $E_u$ and axial-polarized $A_{2u}$ in a $MO_6$ cluster with tetragonal distortion. The metal orbital $3d_{z^2}$ is of symmetry $e_g$ relative to the inversion center at the normal metal site, while the two-oxygen orbital frame vibrates as an $a_{2u}$ bi-orbital. Accordingly, a Pseudo-Jahn-Teller mixing of $e_g$ and $a_{2u}$ by the $A_{2u}$ mode is conceivable at $t_{M-O} = 0$ resulting in the configurational (double-well) instability of the ground-state doublet. As the hybridization is switched on, however, the doublet levels split though lightly as coupling to the $E_g$ mode with axial polarization now results in a Dynamic-Jahn-Teller effect which relaxes the instability at the upper component while leaving a configurational instability at the lower one. The $E_g$ coupling would be allowed by group-theory, in so far as $A_{2u} \otimes A_{2u} = E_g \oplus A_{2g} \oplus \ldots$ . (In the alternative case of an $e_g$ orbital symmetry, the symmetric quadrate $E_g \otimes E_g = E_g \oplus A_{1g}$ decomposes into these same symmetry- breaking and symmetry-retaining components).

As above, a due account of the point-group would require introducing two coordinates, $Q_E$ and $Q_A$, so that the complete interaction DJT Hamiltonian would read:

$$H_{int} = \sum_n g_E (a_{n2}^\dagger a_{n1} + a_{n1}^\dagger a_{n2}) Q_E + \sum_n g_A (a_{n2}^\dagger a_{n2} - a_{n1}^\dagger a_{n1}) Q_A \tag{60}$$

The configurational distortion at the ground-state level is in elongating the axial M(P)-O(A) bonds as observed.

The conclusion that both PJT and DJT couplings constitute the electron-mode interactions at the $MO_6$ octahedra should undoubtedly be taken into account while considering quasi-1D polaron propagation along the c-axis. It appears that the polaron transport along the axis in the cuprates should mainly be controlled by scattering from the ground-state double-well instability [13].

## 8. The mixing phonon coordinate

In order to give the variational equations a more compact form, we rewrite them as

$$\beta_{q\mu}^{\kappa} = -(g^{\mu\nu}/\eta\omega)(\alpha_\nu^\kappa/\alpha_\mu^\kappa) S_{0\kappa}^{\nu\mu}(D_{q\nu\mu}^\kappa/D_q^\kappa) + (g^{\nu\mu}/\eta\omega)^2 S_{0\kappa}^{\nu\mu} Q_{\nu\mu}^\kappa (1/D_q^\kappa) \tag{61}$$

$$\beta_{q\nu}^{\kappa} = -(g^{\nu\mu}/\eta\omega)(\alpha_\mu^\kappa/\alpha_\nu^\kappa) S_{0\kappa}^{\mu\nu}(D_{q\mu\nu}^\kappa/D_q^\kappa) + (g^{\mu\nu}/\eta\omega)^2 S_{0\kappa}^{\mu\nu} Q_{\mu\nu}^\kappa (1/D_q^\kappa) \tag{62}$$

From there and the definition of a phonon coordinate, we get

$$Q_{\nu\mu}^\kappa = -(g^{\mu\nu}/\eta\omega)(\alpha_\nu^\kappa/\alpha_\mu^\kappa) S_{0\kappa}^{\nu\mu}(1/N)\sum_q (D_{q\nu\mu}^\kappa/D_q^\kappa) + (g^{\nu\mu}/\eta\omega)^2 |S_{0\kappa}^{\mu\nu}|^2 Q_{\nu\mu}^\kappa \times$$

$$(1/N)\sum_q(1/D_q^\kappa) - (g^{\nu\mu}/\eta\omega)(\alpha_\mu^\kappa/\alpha_\nu^\kappa)^* S_{0\kappa}^{\mu\nu*}(1/N)\sum_q(D_{-q\,\mu\nu}^\kappa/D_{-q}^\kappa) +$$

$$(g^{\mu\nu}/\eta\omega)^2 |S_{0\kappa}^{\mu\nu}|^2 Q_{\mu\nu}^{\kappa*}(1/N)\sum_q(1/D_{-q}^\kappa) \tag{63}$$

In as much as the D's are all real, if the phonon amplitudes are in-phase within the two component bands, $S_{0\kappa}^{\mu\nu}$ is real too and the variational equations readily split into two equations for the real and imaginary parts, respectively, according to:

$$|Q_{\mu\nu}^\kappa| = \{-(g^{\mu\nu}/\eta\omega) S_{0\kappa}^{\mu\nu}[|(\alpha_\nu^\kappa/\alpha_\mu^\kappa)|(1/N)\sum_q(D_{q\,\nu\mu}^\kappa/D_q^\kappa) +$$

$$|\alpha_\mu^\kappa/\alpha_\nu^\kappa|(1/N)\sum_q(D_{-q\,\mu\nu}^\kappa/D_{-q}^\kappa)\}\exp(i\,2\,\phi_\alpha)/$$

$$\{\exp(i\,\phi_Q)-\exp(-i\,\phi_Q)(g^{\mu\nu}/\eta\omega)^2(S_{0\kappa}^{\mu\nu})^2(1/N)\sum_q[(1/D_q^\kappa)+(1/D_{-q}^\kappa)]\} \tag{64}$$

We have set $\alpha_\mu^\kappa = |\alpha_\mu^\kappa|\exp(+\phi_\alpha)|$, $\alpha_\nu^\kappa = |\alpha_\nu^\kappa|\exp(+\phi_\alpha)|$, $Q_{\mu\nu}^\kappa = |Q_{\mu\nu}^\kappa|\exp(-\phi_Q)$, $Q_{\nu\mu}^\kappa = Q_{\mu\nu}^{\kappa*}$.

The complex $Q_{\mu\nu}^\kappa$ form a specific basis for the irreducible representation $\Sigma_u$ of the $D_{\infty h}$ point group of the infinite linear chain. Another basis for the same representation is built up by the real phonon coordinates $Q_{\mu\nu}^\kappa$. The latter obtain from the above equation setting $\phi_Q = 0, \pi$ which, however, leads automatically to $\phi_\alpha = 0, \frac{1}{2}\pi$. (The latter case of purely imaginary $\alpha$'s would not bring in any novel physics).

We also see that if $Q_{\mu\nu}^\kappa$ are real, so are all the remaining quantities entering in the variational equations, provided the phase $\phi_q^\kappa$ of the phonon amplitude $\beta_{q\mu}^\kappa$ is independent of the band label, as assumed. We remind that under the latter condition all band-to-band pairs of respective phonon components in momentum space propagate in phase to make the off-diagonal DW factor real. Under these conditions, the variational vibronic polaron problem can be solved in real functions.

The integrals incorporated in the above equations should be evaluated analytically if this would give the variational equations a more compact form before the numerical stage was initiated.

Finally, we summarize the resulting equations for solving our variational problem in real functions:

$$Q_{\mu\nu}^{\kappa} = \pm \{ - (g^{\mu\nu}/\eta\omega) S_{0\kappa}^{\mu\nu} [(\alpha_{\nu}^{\kappa}/\alpha_{\mu}^{\kappa}) I_{+1\nu\mu}^{\kappa} + (\alpha_{\mu}^{\kappa}/\alpha_{\nu}^{\kappa}) L_{1\mu\nu}^{\kappa} ] \} /$$

$$\{1 - (g^{\mu\nu}/\eta\omega)^{2} (S_{0\kappa}^{\mu\nu})^{2} (I_{+0}^{\kappa} + L_{0}^{\kappa})\} \qquad (65)$$

where the integrals are:

$$I_{+1\nu\mu}^{\kappa} = (1/N) \sum_{q} (D_{+q\nu\mu}^{\kappa} / D_{+q}^{\kappa}) \qquad (66)$$

$$L_{1\mu\nu}^{\kappa} = (1/N) \sum_{q} (D_{-q\mu\nu}^{\kappa} / D_{-q}^{\kappa}) \qquad (67)$$

$$I_{+0}^{\kappa} = (1/N) \sum_{q} (1 / D_{+q}^{\kappa}) \qquad (68)$$

$$L_{0}^{\kappa} = (1/N) \sum_{q} (1 / D_{-q}^{\kappa}) \qquad (69)$$

with

$$D_{q\mu\nu}^{\kappa} = 1 + 4 (j_{\mu} / \eta\omega) S_{\mu\mu}^{\kappa} \sin(\kappa - \Phi_{\mu\mu}^{\kappa} - q/2) \sin(q/2) - (g^{\mu\mu} / \eta\omega)(\alpha_{\nu}^{\kappa} / \alpha_{\mu}^{\kappa}) Q_{\nu\mu}^{\kappa} S_{0\kappa}^{\nu\mu} \qquad (70)$$

$$D_{\pm q}^{\kappa} = D_{\pm q \nu\mu}^{\kappa} D_{\pm q}^{\kappa} - (g^{\mu\nu} / \eta\omega)^{2} (S_{0\kappa}^{\mu\nu})^{2} (Q_{\mu\nu}^{\kappa})^{2} \qquad (71)$$

The remaining quantities, such as DW factors and phases are:

$$S_{0\kappa}^{\mu\nu} = \exp\{-(1/2N) \sum_{q} (|\beta_{q\mu}^{\kappa}| - |\beta_{q\mu}^{\kappa}|)^{2}\} \qquad (72)$$

$$S_{\mu\mu}^{\kappa} = \exp\{-(1/N) \sum_{q} |\beta_{q\mu}^{\kappa}|^{2} [1 - \cos(q)]\} \qquad (73)$$

$$\Phi_{\mu\mu}^{\kappa} = (1/N) \sum_{q} |\beta_{q\mu}^{\kappa}|^{2} \sin(q) \qquad (74)$$

for use in the numerical calculations.

## 9. Conclusion

We extended Merrifield's Variational Ansatz so as to cover the vibronic polaron occuring as two narrow electronic bands are mixed by an Einstein phonon of appropriate symmetry. The limitations imposed by Group Theory were accounted for carefully, as were other important aspects required for applying the method to specific experimental situations. Ultimately we derived variational equations for the phonon amplitudes and for the fractional contribution of either constituent electronic band to the trial variational state.